\documentclass[12pt,preprint]{aastex}
\begin{document}
\title{Are Gamma-ray Bursts Universal?}
\author{David Eichler\altaffilmark{1} \& Amir Levinson \altaffilmark{2}}
\altaffiltext{1}{Physics Department, Ben-Gurion University,
Beer-Sheva 84105, Israel; eichler@bgumail.bgu.ac.il}
\altaffiltext{2}{School of Physics and Astronomy, Tel-Aviv University,
Tel Aviv 69978, Israel; Levinson@wise.tau.ac.il}
\begin{abstract}
It is noted that the Liang-Zhang correlation can be accounted for
with the viewing angle interpretation proposed earlier. The
Ghirlanda correlation, recently generalized by Nava et al (2006)
to a wind profile, can be accounted for by the viewing angle
interpretation accordingly generalized to a wind profile. Most of
the scatter in the spectra and time-integrated brightness in
$\gamma$-ray bursts (GRB) can thus be accounted for by variation in
two parameters, 1) the viewing angle and 2) the jet opening angle,
with very little variation in any other intrinsic parameters. The
scatter in apparent isotropic equivalent fluence and other
parameters is reduced by a factor of order 30 when each of these
parameters is considered. Possible difficulties with alternative
explanations are briefly discussed. It is also noted that the
relative scatter in the Amati and Ghirlanda correlations suggests
certain conclusions about the inner engine.
\end{abstract}


\section{Introduction}
Several years ago Frail et al. (2001) argued that the  $\gamma$-ray
energy $E_{\gamma,j}$ in $\gamma$-ray bursts (GRB) had much less
variation than the isotropic equivalent energy $E_{\gamma,iso}$.
The hypothesis they implied was that the opening angle of the jet,
$\theta$, as determined from the break in the afterglow light
curve, was the major factor in determining the isotropic
equivalent flux. Dimmer GRB, it was concluded, are dimmer because
the same energy is spread out over a larger solid angle. Thus, the
quantity $\theta^2 E_{\gamma,iso}/2$ is the true $\gamma$-ray energy
and this quantity seems to have much less scatter than
$E_{\gamma,iso}$, uncorrected for opening angle.

 It was also noticed by Amati et al that the isotropic equivalent
luminosity $E_{\gamma,iso}$ is to be strongly correlated with the
spectral energy peak $h\nu_{peak}$ as $E_{\gamma,iso}\propto
(h\nu_{peak})^2$. It was pointed out (Eichler \& Levinson 2004) that
the Amati relation is what one would expect if the reduction in
$\nu_{peak}$ was an illusion created by the viewer not being in
the direction of the jet itself. If the viewing offset,
$\Delta=\theta_{obs}-\theta $, is only a small fraction of the jet
opening angle, viz., $\Delta<<\theta$, then the viewer sees
contributions from a solid angle of order $\Delta^2$ and a
spectral energy peak of $h\nu_{peak}=h\nu^{\star}{\cal D}$, where
${\cal D}=(1-\beta\cos\Delta)^{-1}\simeq 2/\Gamma\Delta^2$ is the
Doppler factor of the fluid element closest to the observer. The
Amati correlation then follows because $E_{\gamma,iso}$ is
proportional to $D^2$ (see appendix A in Levinson \& Eichler,
2005). This explanation of the Amati et al relation would not
apply to a pencil beam (Lamb et al., 2004) or a solid filled in
beam (Yamazaki et al. 2004), for it makes the key assumption that
the solid angle of jet that contributes to the observed GRB
luminosity is proportional to the square of the viewing angle
offset.  It is consistent with the observed relative frequencies
of GRB and X-ray flashes only if the jet has a nontrivial geometry
so that a large faction of all viewers are rather close to the
perimeter of the jet.

Subsequently, Ghirlanda et al. (2004) reported that $E_{\gamma,j}$
correlates with $\nu_{peak}$ as $E_{\gamma,j} \propto
\nu_{peak}^{1.5}$. The implication was that the solid angle,
namely the ratio of $E_{\gamma,j}$ to $E_{\gamma,iso},$ is itself
correlated with these two quantities. Levinson \& Eichler (2005)
then noted that  the modest  difference between the Amati et al
relation and the Ghirlanda relation could be accounted for in a
natural way without making any assumption of correlation between
the physical opening angle and the jet energy output. (Because
this difference is modest and comparable to the scatter, we regard
our explanation of it as reasonable but preliminary, pending a
larger data set of GRB with known redshifts, peak energies, and
break times.) The difference between the respective exponents in
each relation is naturally accounted for by the fact that the
estimate of the jet opening angle is itself weakly affected by
viewing angle. Although the inferred jet opening angle
$\theta_{j,inf}$ is only very weakly dependent on $E_{iso}$, as
$\theta_{j,inf} \propto E_{iso}^{-1/8}$, an offset viewer would
nonetheless overestimate $\theta_{j,inf}$ because his offset
viewing angle causes him to underestimate the true fluence
$E_{\gamma,iso}$ of the jet. The overestimate of the solid angle
$\pi \theta_{j,inf}^2$, is therefore proportional to
$E_{\gamma,iso}^{-1/4}$ i.e. to $\nu_{peak}^{-1/2}$,  and this is
precisely the difference in the exponents in the Amati and the
Ghirlanda correlations. (The quantity $E_{iso}$ as it appears in
the expression for the inferred jet opening angle is, in fact, the
kinetic energy of the jet per unit solid angle. It was assumed to
be proportional to $E_{\gamma,iso}$ in the above analyses to
within a constant.) The two observed relations thus provide
confirmation that the conclusions of Frail et al., Amati et al.
and Ghirlanda et al. not only consistent but mutually supporting.
That is to say, the residual scatter in $E_{\gamma,j}$, after
making the correction for opening angle variation, is mostly
accounted for by making a viewing offset angle correction and vice
versa. These two corrections together eliminate most of the
variation in both $E_{iso}$ and $E_{\gamma,j}$ and they even
reduce the scatter in the inferred opening angle $\theta_{j,inf}$.
The fact that applying only one of these two corrections leaves
residual scatter is by no means evidence against its validity. It
merely implies that two separate factors influence the measured
$\gamma$-ray fluence $E_{\gamma,iso}$.

Eichler and Jontof-Hutter (2005) noted that the $\gamma$-efficiency
$\epsilon_{\gamma}$, defined to be the ratio of $\gamma$-ray energy to
baryon kinetic energy (estimated from the X-ray afterglow luminosity 
at an observer time t of 10 hours) to correlate with $\nu_{peak}$ as 
$\epsilon_{\gamma} \propto \nu_{peak}^{3/2}$. This is nearly
the same correlation exponent as in the Ghirlanda relation. The implication
is that as $\nu_{peak}$ is decreased, the observed $\gamma$-ray fluence decreases with $\nu_{peak}$ much
faster than the blast energy. This is easily understood in the
viewing angle interpretation of the Amati/Ghirlanda  correlations,
because the true blast energy does not depend on the viewing
angle. Moreover, an off-set observer should see suppressed
afterglow until the blast has decelerated enough to encompass the
observer in the $1/\Gamma$ emission cone of the blast material
that generates the afterglow. This effect is consistent with
observations that have been interpreted (Eichler 2005; Eichler and
Granot 2006) as delayed afterglow onset such as gaps between the
prompt emission and the apparent beginning of the afterglow
emission (Piro et al. 2005; Nousek et al. 2005 and references
therein).

Explanations of the Amati and Ghirlanda correlations that posit a
true physical dependence of GRB energy on spectral peak where both
vary considerably (e.g. Rees and Meszaros 2005) leave unanswered
the question of why the GRB energy should have a range of several
orders of magnitude while the blast energy over the same data set
 shows a far less noticeable variation. If the $\gamma$-ray
photosphere is controlled by (the electron counterpart of) a
baryonic component, then the blast energy might be  expected to
vary at least as much as the $\gamma$-ray energy, because radiative
energy is transferred to the kinetic energy of the baryons during
the adiabatic expansion below the photosphere. It may, of course,
be that the photosphere is controlled by pairs (e.g. Eichler 1994;
Eichler and Levinson 2000), a possibility seriously considered by
many other authors as well, and that the difference between bright
$\gamma$-ray bursts and dim ones is expressed primarily by the
emission from a pair dominated photosphere, but the question of
how the baryon dependence scales with burst energy would be left
open. \footnote{The objection is sometimes raised that a pair
dominated photosphere could not have a non-thermal photon
spectrum, however, we see no reason why not. See, for example,
Blandford and Payne 1982; Eichler 1994. Moreover, non-thermal
radiation can be generated it the optically thin region
independently of radiation from the photosphere.}

Liang and Zhang (2005) found the following relation:
\begin{equation}
E_{\gamma,iso,52}=(0.85\pm0.21)\left(h\nu_{peak}\over 100
KeV\right)^{1.94\pm0.17}t_{break,d}^{-1.24\pm0.23} ,\label{LZ}
\end{equation}
where $E_{\gamma,iso,52}$ is the observed isotropic equivalent
$\gamma$-ray luminosity in units of $10^{52}$ erg/s, $h\nu_{peak}$ is the
peak energy as usual, and $t_{break,d}$ is the break time of the
afterglow light curve measured in days.  All relevant
quantities are measured in the cosmological rest frame.  A similar relation
was found later by Nava et al. (2006) using a different method.
As pointed out by Nava et al. (2006), both results appear to be
consistent, within the errors,  with
\begin{equation}
E_{\gamma,iso}\propto (h\nu_{peak})^2t_{break}^{-1}.
\label{LZ2}
\end{equation}
This expresses  a scatter in the Amati relation that follows from
the scatter in the inverse break time.  Note that $E_{\gamma,iso}$
is less for an offset observer than the "true" $E_{iso}$ for an
observer in the beam. Below, we offer a simple explanation for
this relation.

Consider a conical jet of kinetic energy $E_j$, Lorentz factor
$\Gamma$ and semi-opening angle $\theta$, expanding into an
external medium of density $n(r)=\kappa r^{-d}$, where $\kappa$ is
some constant and $r$ is the distance from the center of the
explosion.  In the adiabatic regime the total energy is conserved,
and the evolution of the gas behind the forward shock is given by
(e.g., Meszaros et al. 1998)
\begin{equation}
E_{j}\propto \kappa\Gamma^{2}r^{3-d}\theta^2.
\end{equation}
In terms of the observer time, $dt=dr/\Gamma^2$, and the jet
isotropic equivalent energy, defined as $E_{iso}=\theta^{-2}E_j$,
we have
\begin{equation}
\Gamma(t)\propto \left(E_{iso}\over \kappa\right)^{1/(8-2d)}
t^{(d-3)/(8-2d)}.
\end{equation}
Let $t_j$ denote the time at which $\theta=\Gamma^{-1}$.  Using
the last equation we obtain
\begin{equation}
\theta\propto \left(E_{iso}\over \kappa\right)^{-1/(8-2d)}
t_j^{(3-d)/(8-2d)},
\end{equation}
and
\begin{equation}
E_j= \theta^2E_{iso}/2 \propto \kappa \left({E_{iso} t_j\over
\kappa}\right)^{(3-d)/(4-d)}. \label{Ej2}
\end{equation}
Now, assume that a fraction $\eta_\gamma$ of the kinetic energy is
emitted as gamma rays. (We allow for the possibility that
$\eta_\gamma$ is greater than 1 and in this regard $E_j$ should be
distinguished from the total energy which is the sum of the
kinetic and radiative energy.) The observed isotropic $\gamma$-ray
energy measured by an observer observing the source at some
viewing angle outside the jet that corresponds to an observed peak
energy $h\nu_{peak}$ is $ E_{\gamma,iso}\propto \eta_\gamma
E_{iso}(\nu_{peak}/\nu^\star)^2$, where $h\nu^\star$ defines the
spectral peak energy that will be measured by an on-axis observer
(Eichler \& Levinson 2004; Levinson \& Eichler 2005). By employing
eq. (\ref{Ej2}) to eliminate $E_{iso}$, we finally arrive at:
\begin{equation}
E_{\gamma,iso}\propto \eta_\gamma E_{iso}(\nu_{peak}/\nu^\star)
^2\propto \eta_\gamma \kappa^{1/(d-3)}
E_{j}^{(4-d)/(3-d)}t_{j}^{-1}(\nu_{peak}/\nu^\star)^2. \label{Egiso}
\end{equation}
By associating the observed break time of the afterglow emission
with $t_j$ as commonly done, viz., $t_{break}=t_j$, and assuming
$t_{break}$ to be independent of viewing angle (as is the case
when the observer is within the $1/\Gamma$ emission cone of the
afterglow by the time of the break), we conclude that relation
(\ref{Egiso}) is consistent with the Liang/Zhang relation, as
given in eq. (\ref{LZ2}), provided the quantity $\eta_\gamma
\kappa^{1/(d-3)} E_{j}^{(4-d)/(3-d)}$ is universal. For $d=0$,
this is close to stipulating  that $\eta_{\gamma}E_j=E_{\gamma,j}$
is universal.

Next, consider the collimation corrected energy.  The jet opening
angle $\theta_{j,inf}$ inferred by an off-axis observer that
measures isotropic $\gamma$-ray energy $E_{\gamma,iso}$ and break
time $t_{break}$, and who assumes ambient medium with density
profile as above, satisfies
\begin{equation}
\theta_{j,inf}\propto \left(E_{\gamma,iso}\over
\kappa\right)^{-1/(8-2d)}t_{break}^{(3-d)/(8-2d)}.
\end{equation}
The collimation corrected energy that will be obtained by using
the latter expression for the jet semi-opening angle is then
\begin{equation}
E_{\gamma,inf}=\theta_{j,inf}^2 E_{\gamma,iso}\propto
\eta_\gamma^{(5-d)/4-d)}E_{j}(\nu_{peak}/\nu^\star)^{(6-2d)/(4-d)},
\end{equation}
which, for a universal $\eta_\gamma^{(5-d)/4-d)}E_j$, is
consistent with the Ghirlanda relations for both a uniform density
medium (d=0) and a wind profile (d=2), as discussed in Nava et al.
(2006).  This is not surprising, since  the connection between
$E_{\gamma, iso}$ and $E_{\gamma,inf}$ is defined in Nava et al
(2006) in the same way as here. Regardless of what one assumes
about the surrounding density profile, the  point remains (as
already noted by Nava et al with  different phraseology) that the
Amati correlation and the Frail correlation between
$E_{\gamma,iso}$ and $\theta_{j,inf}^{-2}$ imply the Liang and
Zhang correlation given the standard assumptions of afterglow
theory. Moreover, the equivalence of the Amati correlation and
Ghirlanda correlation    if the former is attributed to viewing
angle effects (Levinson and Eichler, 2005) is independent of
assumptions about the surrounding density profile.
 The small scatter in the Ghirlanda
and Liang/Zhang relations indicates that the ``true'' GRB energy
is universal, as claimed originally by Frail et al (2001).

\section {Conclusions and Further Discussion}

As noted by Nava et al (2006), the Ghirlanda correlation and its
physical implications change with assumption about the surrounding
density profile. This is because, unlike the Amati correlation,
the Ghirlanda correlation is not one of purely observed
quantities, but rather includes within it a theoretical inference
about the GRB jet opening angle that depends on assumptions
regarding the evolution of the blast wave. In particular, they
note that if they assume a wind-like profile, (together with the
tacit assumption that opening angle is uncorrelated with
$E_{\gamma,j}$) then the GRB energy $E_{\gamma,j}$ scales linearly
with $h\nu_{peak}$, and the photon entropy is constant among the
different bursts, whereas this conclusion would not follow if a
constant ambient density profile were assumed. The question would
remain open as to why the photon entropy would remain constant
over a wide range of $E_{\gamma,j}$ and $h\nu_{peak}$, especially
if the latter is established at a pair-dominated photosphere.

Here we have shown that the viewing angle interpretation of both
the Amati and Ghirlanda correlations, and the equivalence between
the two is independent of assumptions about the ambient density
profile. This is because the universality of the GRB energy $E_j$
implied by this equivalence is a physically separate issue from
the opening angle ( the latter presumably established by
collimation well downstream of the central engine), and is
therefore unaffected by it. A set of GRBs with identical $E_j$
could be placed in an environment of any density profile and the
theoretical values of $E_j$, if correctly inferred by making the
correct assumptions about the density profile, would all yield the
same conclusion - that the range of $E_j$ is narrow.

That the Ghirlanda relation shows less scatter than the Amati
correlation is significant in the same way that the Frail
correlation (for a limited range of spectral peak) is. We
interpret it to mean that modest variation in the opening angle
introduces additional scatter into the observed $E_{\gamma,iso}$
after either $E_{\gamma,j}$ or  $E_{j}$ has been established by
the central engine. This is to be contrasted with the reverse
situation: that $E_{iso}$ in an outflow is established by the
central engine and the $\gamma$- ray output $E_{\gamma,j}$ is
established, say, by internal shocks whose effective covering
solid angle or overall efficiency varies from one GRB to the next.
In the latter case, one would expect more scatter in
$E_{\gamma,j}$ than in $E_{iso}$ due to the  additional scatter in
the covering angle. The low scatter $E_{\gamma,j}$ is consistent
with, and perhaps even supportive of the claim (Eichler and
Jontof-Hutter 2005, Eichler and Granot 2006) that the $\gamma$-ray
efficiency is close to 100 percent in GRBs, and that only a small
fraction of the energy is in blast energy, since this is a
reliable way of limiting the $\gamma$-ray efficiency to a narrow
range..

If  the Ghirlanda correlation indeed proves to have  a different
slope from the Amati correlation this will also be significant. It
 would imply that the inferred opening angle varies systematically
with $h\nu_{peak}$. At present, the implied systematic variation
is only comparable to the scatter in solid angle inferred from the
observed  break times. If there were a wide range of physical jet
energies and true spectral peaks, the difference in slope would
then be a considerable spread in opening angles associated with
the wide range of $h\nu_{peak}$. Specifically, if the range of
$h\nu_{peak}$ is from 30 KeV to 1 MeV, and, as assumed by Nava et
al (2006), the ambient density is wind-like, then the Ghirlanda
correlation would be $E_j \propto h\nu_{peak}$ and it would then
follow that $\theta_{inf}^{-2} \propto h\nu_{peak}$. It would
follow that the range of solid angles is about 30, scaling in
inverse proportion to $h\nu_{peak}$.

 We also note that even though
$E_{iso}$ does not appear to  have much remaining scatter after
the various correlations discussed here are accounted for,
$L_{iso}$, the isotropic equivalent luminosity does, because the
durations of long bursts vary from several seconds to several
hundreds of seconds. Any physical mechanism that ties $E_{iso}$ to
$h\nu_{peak}$ would have to tolerate the large variation in GRB
duration and the attendant variation in $L_{iso}$ for a given
$E_{iso}$. This is significant because the bulk Lorentz factor at
the photosphere, which is likely to enter into $h\nu_{peak}$ in
some models, is more likely to depend on $L_{iso}$ than on
$E_{iso}$.

 Yet another significant statistic, in our view, is that the blast energy
 does {\it not} correlate
nearly as noticeably with $h\nu_{peak}$ as does $E_{\gamma,inf}$.
In fact, the best fit for the ratio $E_{\gamma,inf}/E_{k}$ has it
correlating linearly as $h\nu_{peak}^{1.4}$ (Eichler and
Jontof-Hutter, 2005) which is nearly exactly the Ghirlanda
relation [Here $E_{\gamma,inf} $ is the inferred $\gamma$-ray output
and $E_{k}$ is the kinetic energy of the ejected mass as inferred
from the 10 hour X-ray afterglow (Freedman and Waxman 2001,
Lloyd-Rhonning and Zhang 2004). Note that $E_k$ is often used
interchangeably with the quantity $E_j$ as defined in Equation 3.]
If the Amati correlation were to be attributed to
real physical variations in both $E_{iso}$ and $h\nu_{peak}$ that
are closely tied together, then it would suggest that bright, hard
GRB are brighter than dim, soft ones not primarily because of more
baryon kinetic energy, but rather because of greater dominance of
other forms of energy. Presumably, the non-baryonic energy is
mostly photons and pairs; the point is that it generates more
photons without generating noticeably more afterglow. This would
be consistent with the best estimate of $E_{\gamma,inf}/E_k$ for
the brightest bursts that is considerably greater than unity
(Eichler and Jontof-Hutter 2005, Eichler and Granot 2006).
However, in the simplest model of an adiabatically expanding
baryon-free fireball that expands from a fixed dissipation radius
$R_0$, the photon entropy is proportional to $E_{tot}^{3/4}$ where
$E_{tot}$ is the total energy. So the variation of $R_0$ with
$E_{tot}$ would have to be tailored to obtain a fit with the
Ghirlanda correlation, which, for a wind-like ambient density
profile, gives a constant photon entropy.

To conclude, when both the $h\nu_{peak}$ and $t_{break}$
correlations with $E_{iso}$ are accounted for, the remaining
scatter in the latter quantity is remarkably small, less than a
factor of 2 (e.g. Nava et al, 2006). This suggests that some
quantity in GRBs is universal. The pure viewing angle
interpretation of the Amati correlation posits that it is both the
jet energy and spectral peak that vary little from one GRB to the
next, while modest random variation in opening angle is acceptable
and systematic variation of the opening angle with
$E_{\gamma,iso}$ is not implied. The most natural underlying
explanation of why this should be the case, we suggest, is that
baryonic contamination is too small to affect the quantity of
primary $\gamma$-ray emission (e.g. Levinson \& Eichler, 1993;
Eichler \& Levinson 2000), and that, therefore, neither should the efficiency of
internal shocks affect the overall energy output in $\gamma$-rays.
By contrast, the non-thermal component of the prompt $\gamma$-ray
emission, which does indeed vary considerably among GRBs, may well
depend on such factors. Similarly, the details of the erratic
behavior of the light curve, in which there is considerably
variety, may well depend on the less predictable aspects of GRB
such as the internal shocks and baryon contamination.

This research was supported by the Israel-US Binational Science Foundation,
an Israel Science Foundation Center of Excellence Award, and the Arnow Chair of Theoretical Physics.


\begin{thebibliography}{99}


\bibitem[] {1} Amati, L. et al., 2002, A\&A, 390, 81
\bibitem[] {2} Blandford, R.D. \& Pyne, D.G. 1982, MNRAS, 199, 883
\bibitem[] {7} Eichler, D. 1994, ApJS, 90, 877
\bibitem[] {8} Eichler, D. \& Granot, J. 2006, ApJ, in press (astro-ph/0509857)
\bibitem[] {8} Eichler, D. \& Jontof-Hutter, D. 2005, ApJ, 635, 1182
\bibitem[] {9} Eichler, D. \& Levinson A., 2000, ApJ, 529, 146
\bibitem[] {10} Eichler, D. \& Levinson A., 2004, ApJ, 614, L13
\bibitem[] {11} Frail, D. A. et al., 2001, Ap. J. 562, L55
\bibitem[] {12} Freedman, D.L., and Waxman, E., 2001 ApJ, 547, 922
\bibitem[] {13} Ghirlanda, G., Ghisellini, G. \& Lazzati, D. 2004, ApJ, 616, 331
\bibitem[] {14} Lamb, D.Q., et al. 2004, NewA Rev., 48, 459
\bibitem[] {15} Levinson, A. \& Eichler, D. 1993, ApJ, 418, 386
\bibitem[] {16} Levinson, A. \& Eichler, D. 2005, ApJ, 629, L13
\bibitem[] {17} Liang, E. \& Zhang B. 2005, ApJ, 633, 611
\bibitem[] {18} Lloyd-Ronning, N., and Zhang, B., 2004, ApJ, 613, 477

\bibitem[] {19} Meszaros, P. Rees, M.J., \& Wijers, R.A.M.J. 1998, 499, 301

\bibitem[] {20} Nava, L. et al. 2006, A\&A, in press
\bibitem[] {21} Nousek, J.A., et al. 2005, submitted to ApJ (astro-ph/0508332)
\bibitem[] {22} Piro, L., et al. 2005, ApJ, 623, 314
\bibitem[] {23} Rees, M.J. \& Meszaros, P. 2005, ApJ, 628, 847
\bibitem[] {24} Yamazaki, R. et al., 2004, ApJ 606, L33

\end{thebibliography}
\end{document}